\documentclass[a4paper]{jpconf}
\usepackage{graphicx}
\begin{document}
\title{Search for the CP violating $K_S \to 3\pi^0$ decay with the KLOE detector}

\author{M. Silarski on behalf of the KLOE--2 Collaboration\footnote[1]{The KLOE--2 Collaboration:
D.~Babusci, D.~Badoni, I.~Balwierz-Pytko, G.~Bencivenni, C.~Bini, C.~Bloise, F.~Bossi, P.~Branchini, 
A.~Budano, L.~Caldeira~Balkest\aa hl, G.~Capon, F.~Ceradini, P.~Ciambrone, F.~Curciarello, E.~Czerwi\'nski, E.~Dan\`e,
V.~De~Leo, E.~De~Lucia, G.~De~Robertis, A.~De~Santis, A.~Di~Domenico, C.~Di~Donato, R.~Di~Salvo, D.~Domenici,
O.~Erriquez, G.~Fanizzi, A.~Fantini, G.~Felici, S.~Fiore, P.~Franzini, P.~Gauzzi, G.~Giardina,
S.~Giovannella, F.~Gonnella, E.~Graziani, F.~Happacher, L.~Heijkenskj\"old, B.~H\"oistad,
L.~Iafolla, M.~Jacewicz, T.~Johansson, K.~Kacprzak, A.~Kupsc, J.~Lee-Franzini, B.~Leverington, F.~Loddo,
S.~Loffredo, G.~Mandaglio, M.~Martemianov, M.~Martini, M.~Mascolo, R.~Messi, S.~Miscetti, G.~Morello,
D.~Moricciani, P.~Moskal, F.~Nguyen, A.~Passeri, V.~Patera, I.~Prado~Longhi, A.~Ranieri, C.~F.~Redmer,
P.~Santangelo, I.~Sarra, M.~Schioppa, B.~Sciascia, M.~Silarski, C.~Taccini, L.~Tortora, G.~Venanzoni,
W.~Wi\'slicki, M.~Wolke, J.~Zdebik
}}

\address{Institute of Physics, Jagiellonian University, PL-30-059 Cracow, Poland}

\ead{Michal.Silarski@lnf.infn.it}

\begin{abstract}
We present a new search for the $K_S \to 3\pi^0$ decay performed with the KLOE detector operating at the
DA$\Phi$NE $\phi$--factory. The $K_S$ mesons were tagged via registration of $K_L$ mesons
which crossed the drift chamber without decaying and interacted with the KLOE electromagnetic
calorimeter. The $K_S \to 3\pi^0$ decay was then searched requiring six prompt photons.
To suppress background, originating from fake $K_S$ tags and $K_S \to 2\pi^0$ decays with additional
two spurious clusters, we have performed a discriminant analysis based on kinematical fit,
testing of the signal and background hypotheses and exploiting of the differences in kinematics
of the $K_S$ decays into $2\pi^0$ and $3\pi^0$. In a sample of about 1.7$\cdot 10^9$ $\phi \to K_S K_L$
events we have found no candidates in data and simulated background samples.
Normalizing to the number of $K_S \to 2\pi^0$ events in the same sample, we have set the upper
limit on the $K_S \to 3\pi^0$ branching ratio BR($K_S \to 3\pi^0) < 2.6 \cdot 10^{-8}$ at 90\% C.L.,
five times lower than the previous limit. This upper limit
can be translated into a limit on the modulus of the $\eta_{000}$ parameter amounting to $|\eta_{000}| < 0.0088$
at 90\% C.L., improving by a factor two the latest direct measurement.
\end{abstract}

\section{Introduction}
Since the first discovery of the $\mathcal{CP}$--violating neutral kaon decay in 1964, there
has been made a big effort to describe the $\mathcal{CP}$ symmetry breaking within the Standard Model.
The favoured theoretical framework was provided in 1973 by Kobayashi and Maskawa, who pointed out that
$\mathcal{CP}$ violation would follow automatically if there were at least six quark flavours.
At present the main experimental effort is focused on the neutral $B$ and $D$ meson system
studies. However, there are still several interesting open issuses in the kaon physics, which can
contribute to our better understanding of the $\mathcal{CP}$ violation mechanism~\cite{silarskiPHD}.
For example, $K_S$ decays to $|\pi^+ \pi^- \pi^0\rangle$ and $|3\pi^0\rangle$,
requiring $\mathcal{CP}$ violation, are still poorly known.
The $|\pi^+ \pi^- \pi^0\rangle$ final state can be produced in neutral kaon decays with
isospin $I = 0, 1, 2,$ or 3. The $I = 0$ and $I = 2$ states have $\mathcal{CP} = 1$,
and $K_S$ can decay into them without violation of the $\mathcal{CP}$ symmetry. However,
they are expected to be strongly suppressed by centrifugal barrier effects~\cite{pdg2012}.
For the $I = 1$ and $I = 3$ states there is no centrifugal barrier and $\mathcal{CP} = -1$, so $K_S$
decay requires violation of this symmetry. Anyhow the two kinds of final states can be separated
by the analysis of the $\pi^+ \pi^- \pi^0$ Dalitz plot. This allows for determination of the $K_S$ to $K_L$
decay amplitude ratio
$\eta_{+-0} = \frac{A( K_{S}\rightarrow \pi^{+}\pi^{-} \pi^0)}{A(K_{L}\rightarrow \pi^{+}\pi^{-}\pi^0)}
\cong \epsilon + \epsilon'_{+-0}$, where $\epsilon$ indicates the $K_S$ $\mathcal{CP}$ impurity and
$\epsilon^{'}_{+-0}$ is the contribution of the direct $\mathcal{CP}$--violating term. 
In the case of $|\pi^0 \pi^0 \pi^0\rangle$ final state, only isospin $I = 1$ or $I = 3$ is allowed,
for which $\mathcal{CP} = -1$. Therefore, the $K_S \to 3\pi^0$
decay is a purely $\mathcal{CP}$ violating process, for which one defines the analogous ratio:
$\eta_{000} = \frac{A( K_{S}\rightarrow \pi^0\pi^0 \pi^0)}{A(K_{L}\rightarrow \pi^0\pi^0 \pi^0)}
\cong \epsilon + \epsilon'_{000}$.\\
The present knowledge about $\eta_{+-0}$ and $\eta_{000}$ is poor mainly due to very low decay
rates for the $K_S \to 3\pi$ decays. The current value of the $K_S \to \pi^+ \pi^- \pi^0$ branching
ratio amounts to $BR(K_S \to \pi^+ \pi^- \pi^0) = (~3.5^{+1.1}_{-0.9}~)\cdot 10^{-7}$~\cite{pdg2012}, and
the $K_S \to 3\pi^0$ has been never observed. The best upper limit on this decay
branching ratio amounts to $BR(K_S \to 3\pi^0) < 1.2 \cdot 10^{-7}$~\cite{Matteo},
while the prediction based on Standard Model is equal to about $2\cdot 10^{-9}$~\cite{MPaver}.\\
In this article we briefly describe the search of the $K_S \to 3\pi^0$ decay based on the 1.7~fb$^{-1}$
of integrated luminosity gathered with the KLOE detector operating at the $\phi$--factory
DA$\Phi$NE of the Frascati Laboratory~\cite{Michal}.
\section{The KLOE detector}
\begin{figure}
\begin{center}
\includegraphics[width=0.4\textwidth]{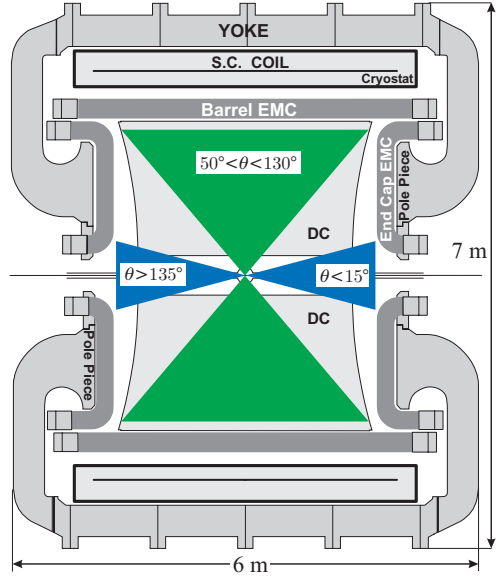}
\caption{Schematic view of the KLOE detector.}
\label{kloe}
\end{center}
\end{figure}
The KLOE experiment operated from 2000 to 2006 at DA$\Phi$NE, the $e^+e^-$ collider
optimized to work with a center of mass energy around
the $\phi$ meson mass peak: $\sqrt{s}$ = 1019.45 MeV~\cite{kloe2008}.
Equal energy positron and electron beams collided at an angle of $\pi - 25$~mrad producing
$\phi$ mesons nearly at rest, which then decayed mainly to $K^+K^-$ (49\%) and $K_SK_L$ (34\%)
final states.
The decay products were registered using the KLOE detection setup presented schematically in Fig.~\ref{kloe}.
The detector consists of a large cylindrical Drift Chamber~\cite{DCH} surrounded by a lead scintillating
fiber Electromagnetic Calorimeter~\cite{EMC}, both immersed in an axial 0.52~T magnetic field produced by a
superconducting solenoid. The beam pipe at the interaction region is a sphere with 10 cm of radius
made of a Beryllium--Aluminum alloy of 0.5 mm thickness.
The drift chamber, 4~m in diameter and 3.3~m long, has 12582 all stereo drift cells with tungsten
sense wires. To minimize the $K_L$ regeneration, multiple Coulomb scattering and photon
absorption the chamber is constructed out of carbon fiber composite with low--Z and low density,
and uses the gas mixture of helium~(90\%) and isobutane (10\%). It provides three--dimensional
tracking with resolution in the bending plane about 200 $\mu$m, resolution on the z--co\-or\-di\-na\-te
measurement of about 2 mm and of 1 mm on the decay vertex position. Momentum of a particle is determined
from the curvature of its trajectory in the magnetic field with a fractional accuracy $\sigma_p/p~=~0.4\%$
for polar angles larger than 45$^{\circ}$~\cite{kloe2008}.
The KLOE electromagnetic calorimeter covers 98\% of the solid angle and is
composed by a barrel  and two endcaps, for a total of 88 modules.
Each module is built out of 1 mm scintillating fibers grouped in cells of 4.4x4.4 cm$^2$
and embedded in 0.5 mm lead foils, and it is read out from both sides by set of photomultipliers.
The particle energy deposits are obtained from the photomultipliers signal amplitude, while the
arrival times and particles impact points are determined from the spatial coordinates of the fired
calorimeter cell and the difference between times measured at both ends of the calorimeter module. 
The accuracies of energy and time determination are parametrized as $\sigma_E/E = 5.7\%/\sqrt{E\ {\rm(GeV)}}$ 
and  $\sigma_t = 57\ {\rm ps}/\sqrt{E\ {\rm(GeV)}} \oplus100\ {\rm ps}$, respectively.
The trigger \cite{TRG} uses both calorimeter and chamber information. In the search for the $K_S \to 3\pi^0$
we have used, however, only the calorimeter information requiring two energy deposits with $E>50$ MeV
for the barrel and $E>150$ MeV for the endcaps.
\section{Data analysis}
\begin{figure}
\begin{center}
\includegraphics[width=0.49\textwidth]{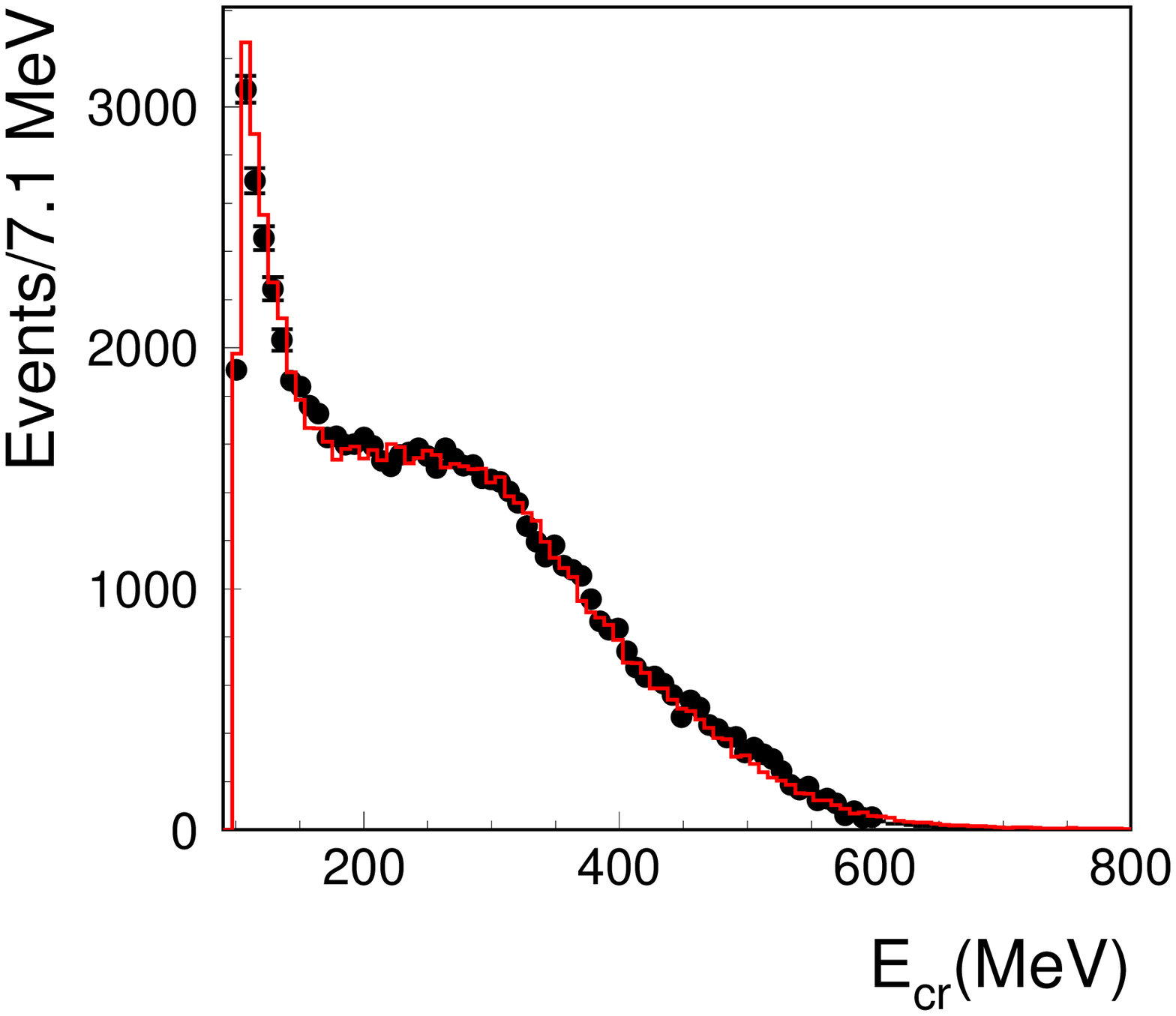}
\includegraphics[width=0.49\textwidth]{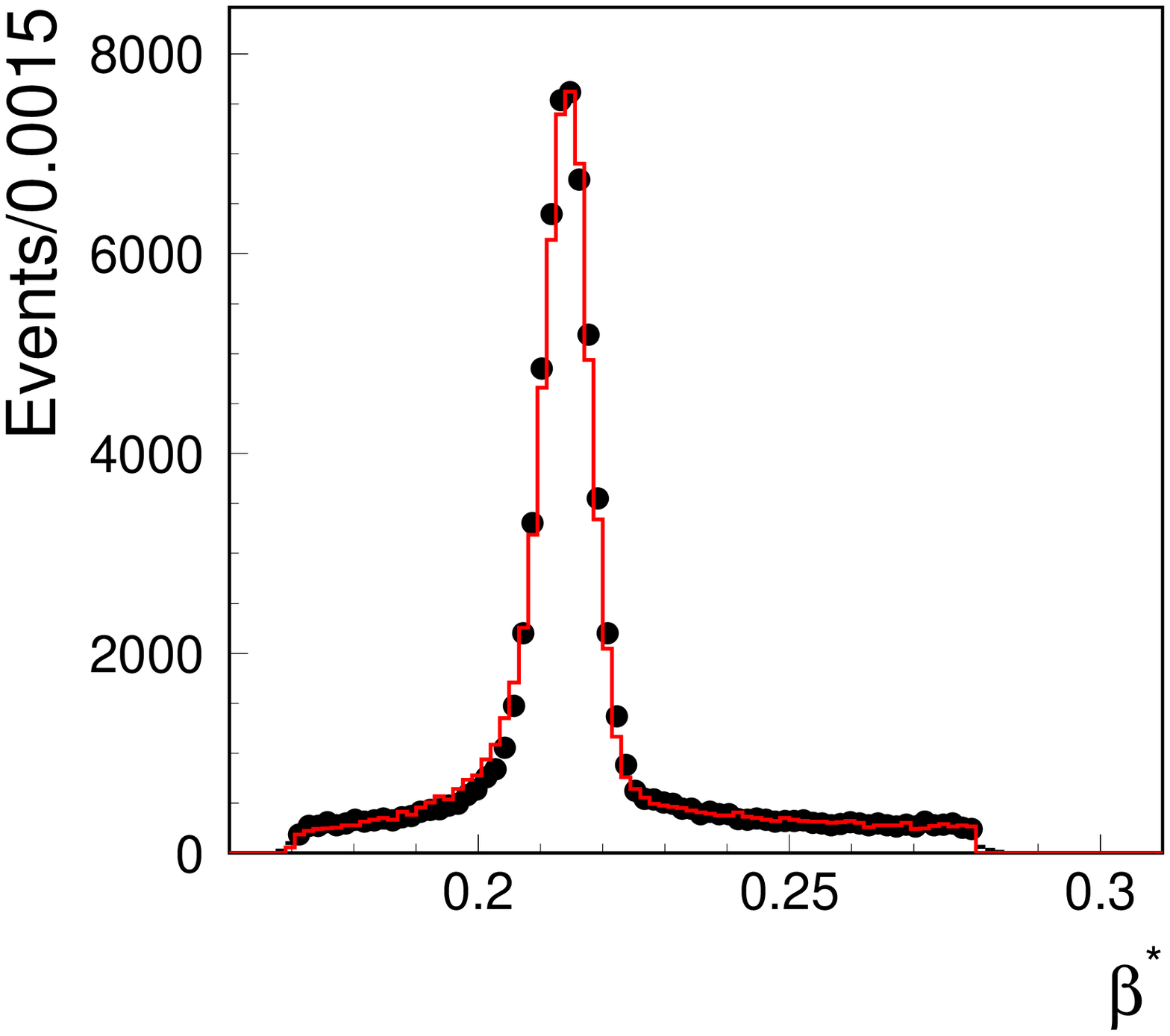}
\end{center}
\caption{Distributions of the reconstructed $K_L$ energy ($E_{cr}$) and velocity
in the $\phi$ center of mass frame ($\beta^*$) for all events in the six photon
sample. Black points represent data, while the MC background simulation is shown
as red histogram.}
\label{fig1}
\end{figure}

\begin{figure}
\includegraphics[width=0.49\textwidth]{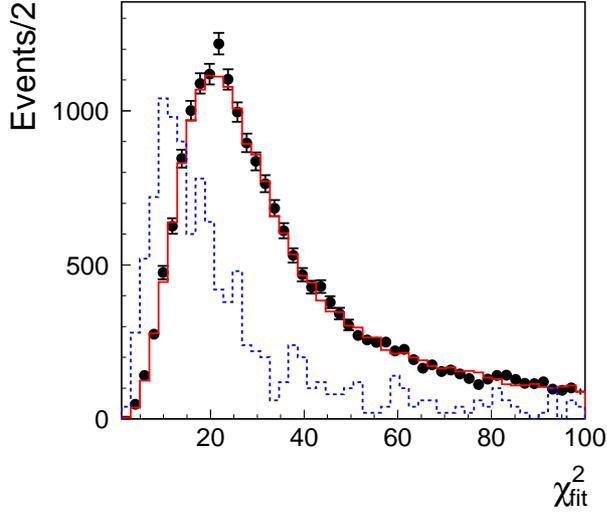}
\hspace{2pc}
\begin{minipage}[b]{16pc}
\caption{\label{label}
Distribution of $\chi^2_{fit}$ for the tagged six--photon sample for data
( black points), background simulation (solid histogram), and simulated $K_S \to 3\pi^0$ signal
(dashed histogram).}
\end{minipage}
\label{fig:chi2}
\end{figure}

\begin{figure}
\begin{center}
\includegraphics[width=0.49\textwidth]{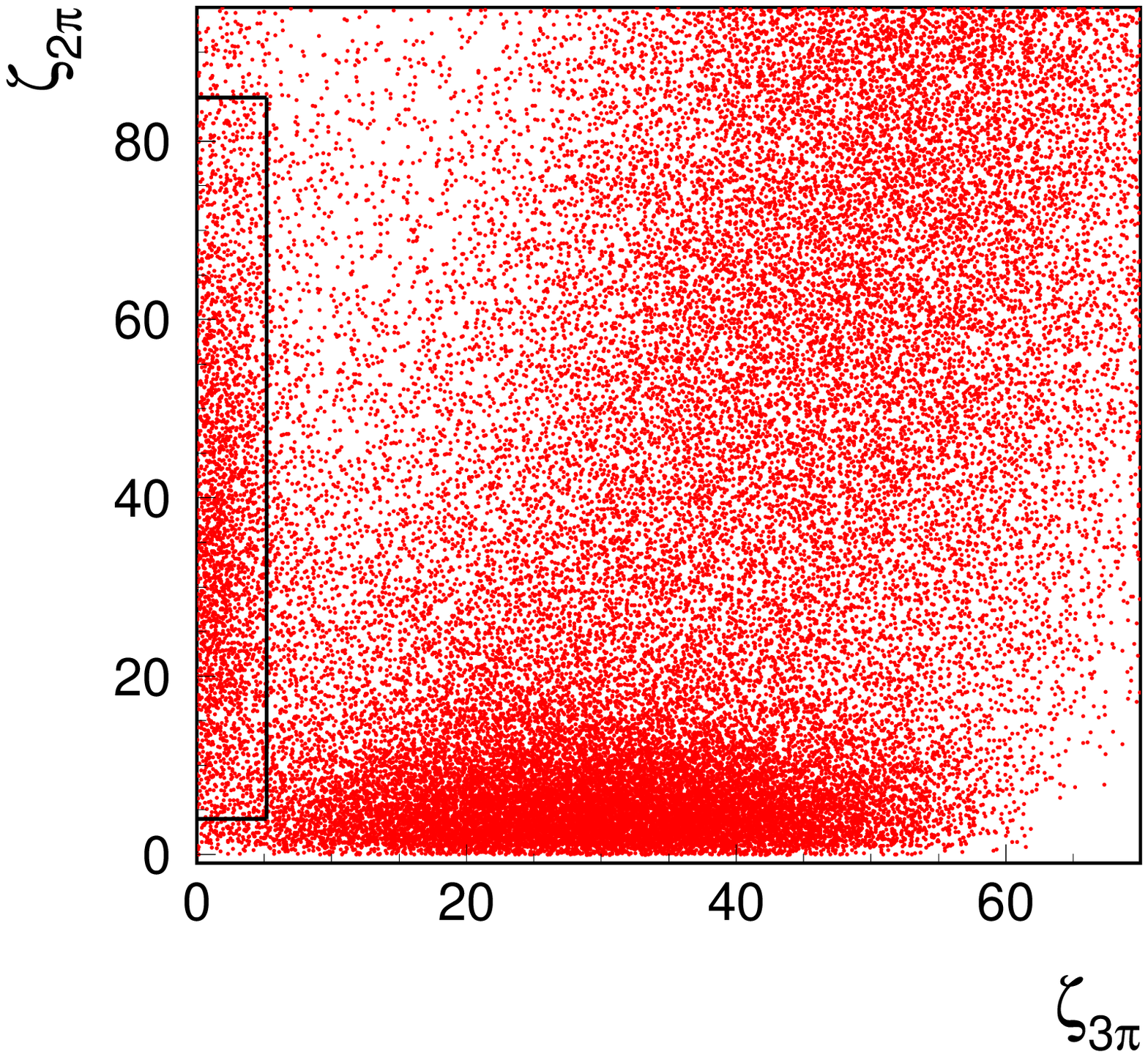}
\includegraphics[width=0.49\textwidth]{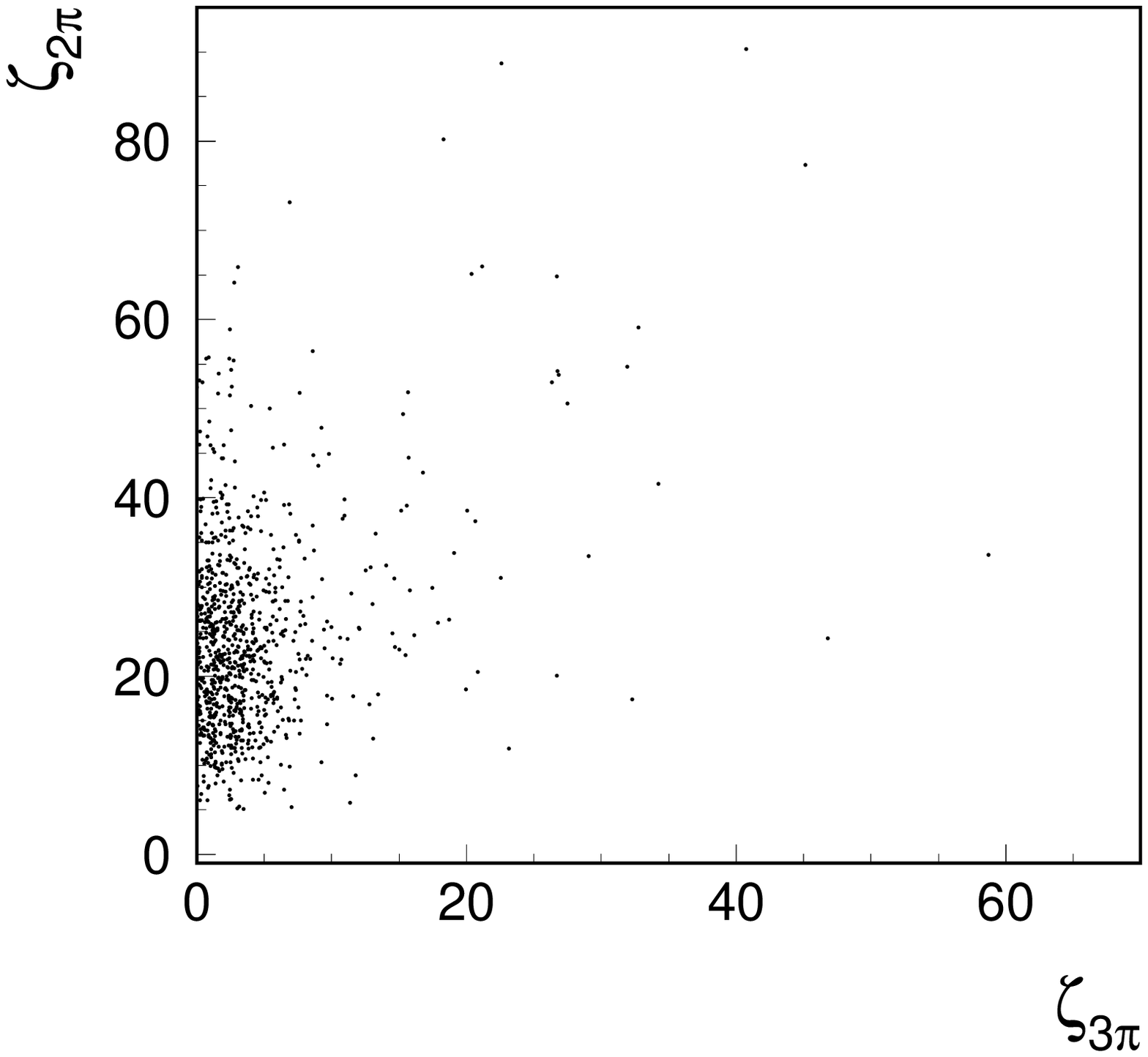}
\end{center}
\caption{Distributions of events in the $\zeta_{3\pi}$-$\zeta_{2\pi}$ plane,
for six--photon sample tagged by $K_L$--crash for data (left),
and for the simulated $K_S\to 3\pi^0$ decays (right).
The solid lines show the signal region definition chosen using the optimization procedure described in the text.}
\label{fig3}
\end{figure}

\begin{figure}
\begin{center}
\includegraphics[width=0.49\textwidth]{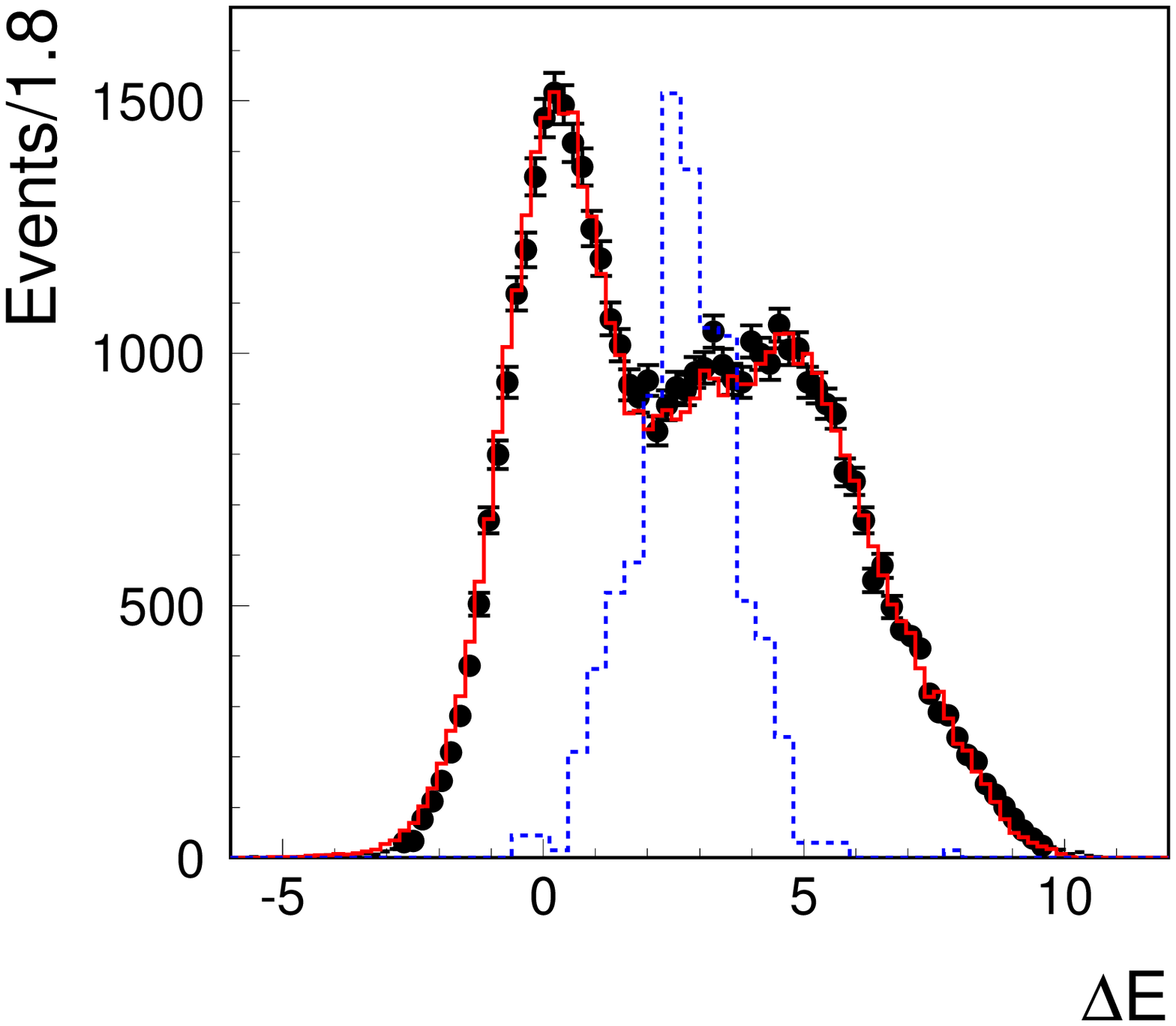}
\includegraphics[width=0.49\textwidth]{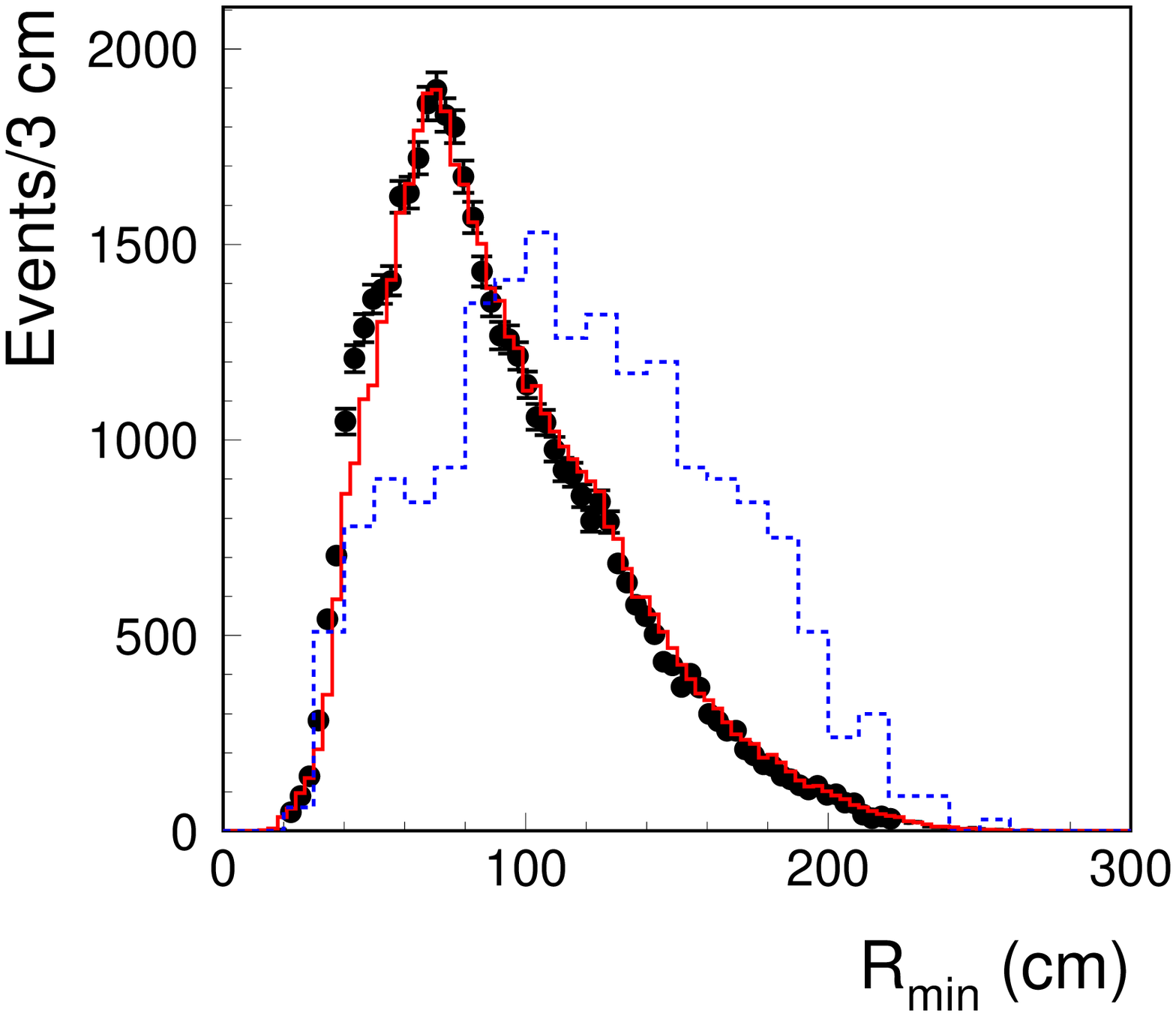}
\end{center}
\caption{Distributions of  $\Delta E$
and $R_{min}$ discriminating variables for six--photon events for data (black points)
and background simulations (red histograms). The dashed histograms represents simulated
$K_S \to 3\pi^0$ events.}
\label{fig5}
\end{figure}

At KLOE kaons arising from the $\phi$ decay move at low speed with their relative angle close to 180$^{\circ}$.
Therefore, observation of a $K_L$ ($K_S$ ) decay ensures the presence of a $K_S$ ($K_L$~)
meson travelling in the opposite direction. About 50\% of $K_L$'s reach the calorimeter before 
decaying, which provides a very clean $K_S$ tag by the $K_L$ interaction in the calorimeter 
($K_L$--crash). It is identified by a cluster with polar angle $40^\circ<\theta_{cr}<140^\circ$,
not associated to any track, with energy $E_{cr}>100$~MeV and with  a time corresponding to a $K_L$
velocity $\beta^*\sim $ 0.2 in the $\phi$ rest frame. The average value of the $e^+e^-$ center of mass 
energy $\sqrt{s}$ is obtained with a precision of 20 keV for each 200 nb$^{-1}$  running period
using large angle Bhabha scattering events~\cite{kloe2008}. The value of $\sqrt{s}$ and the $K_L$--crash
cluster position allows us to obtain, for each event, the direction of the $K_S$ with an angular resolution
of 1$^{\circ}$ and a momentum resolution of about 2 MeV.
The search for the $K_S\to 3\pi^0\to 6\gamma$ decay is then carried out by the selection of events
with six photons which momenta are reconstructed using time and energy measured by the electromagnetic
calorimeter. Photons from the $K_S$ decay are reconstructed as neutral particles that travel with a velocity
$\beta=1$ from the interaction point to the KLOE calorimeter (prompt photons). In order to retain a large control
sample for the background while preserving high efficiency for the signal, we keep all photons satisfying
$E_\gamma>$ 7 ~MeV and $|\cos \theta|<$ 0.915. Each cluster is also required to satisfy the condition 
$|t_{\gamma}-R_{\gamma}/c|<{\rm min}(3.5\sigma_t, 2\ {\rm ns})$, 
where $t_{\gamma}$ is the photon flight time and $R$ the path length.
The photon detection efficiency of the calorimeter amounts to about 90\% for $E_\gamma$ = 20 MeV, and reaches 
100\% above 70 MeV.
To determine the $K_S \to 3\pi^0$ branching ratio we have counted also the $K_S \to 2\pi^0$ events
which are selected requiring 4 prompt photons, and have served as normalization sample.
For both channels the expected background as well as the detector acceptance and the analysis efficiency
are estimated using the Monte Carlo simulation tool of the experiment~\cite{NIMOffline}.
All the processes contributing to
the background were simulated with statistics twice larger than
the data sample. Moreover, for the acceptance and the analysis efficiency evaluation a dedicated
$K_{S}\to 3\pi^{0}$ signal simulation was performed.
After $K_S$ tagging we have counted 76689 events with six prompt photons. For these events a further
discriminant analysis was performed to increase the signal to background ratio.\\
The first analysis step aimed to reject fake $K_S$ tags.
The distributions of reconstructed $K_L$ energy $E_{cr}$ and velocity $\beta^*$ for the selected data sample
and background simulations are shown in Fig.~\ref{fig1}.
In the $\beta^*$ distribution,
the peak around 0.215 corresponds to genuine $K_L$ interactions in the calorimeter,
while the flat distribution mainly originates from
$\phi \to K_{S}K_{L} \to (K_S \to \pi^+\pi^-, K_L \to 3\pi^0)$ background events.
In this case one of the low momentum charged pions spirals in the forward direction and interacts
in the low--$\beta$ insertion
quadrupoles. This interaction produces neutral particles which simulate the signal of $K_L$
interaction in the calorimeter (fake $K_L$--crash), while the $K_L$ meson decays close
enough to the interaction point to produce six prompt photons~\cite{Michal}.
To suppress this kind of background we first reject events
with charged particles coming from the vicinity of the interaction region.
Next, taking advantage of the differences in the $\beta^*$ and $E_{cr}$ distributions between
the tagged $K_S$ events and the fake $K_L$--crash, we have tightened
cuts on these variables: $E_{cr} > 150~\mathrm{MeV}$ and $0.20 <\beta^* < 0.225$.
This improves by a factor 12 the rejection of this background with respect to the previous
analysis~\cite{Matteo}.
The second source of background originates from wrongly reconstructed $K_S\to 2\pi^0$ decays.
The four photons from this decay can be reconstructed as six due to fragmentation of the electromagnetic
showers (splitting). These events are characterized by one or two low--energy clusters reconstructed very close to
the position of the genuine photon interaction in the calorimeter. Additional clusters can come from
the accidental time coincidence between the $\phi$ decay and machine background photons
from DA$\Phi$NE~\cite{Michal}.
In the next stage of the analysis we select only kinematically well defined events.
To this end we perform the kinematical fit procedure based on the least squares method
imposing energy and momentum conservation, the kaon mass and the velocity of the six photons
in the final state. The $\chi^2$ distribution of the fit for data and background simulation,
$\chi^2_{fit}$, is shown in Fig.~\ref{fig:chi2} together with the expected distribution
for signal events. Cutting on $\chi^2_{fit}$ considerably reduces the background 
from bad quality reconstructed events while keeping large signal efficiency~\cite{Michal}.
In order to reject events with split and accidental clusters we look at the correlation between
two $\chi^2$--like discriminating variables $\zeta_{2\pi}$ and $\zeta_{3\pi}$. 
$\zeta_{2\pi}$ is calculated by an algorithm selecting four out of six clusters best satisfying
the kinematic constraints of the two--body decay, therefore it verifies the $K_S \to 2\pi^0 \to 4\gamma$
hypothesis. The pairing of clusters is based on the requirement $m_{\gamma\gamma} = m_{\pi^0}$,
and on the opening angle of the reconstructed pions trajectories in the $K_S$ center of mass frame.
Moreover, we check the consistency of the energy and momentum conservation in the
$\phi \to K_S K_L, K_S \to 2\pi^0$ decay hypothesis.
The $\chi^{2}_{3\pi}$ instead verifies the signal hypothesis by looking at
the reconstructed masses of three pions. For every choice of cluster pairs
we calculate the quadratic sum of residuals between the nominal $\pi^0$
mass and the invariant masses of three photon pairs.
As the best combination of cluster pairs, we take the configuration minimizing $\zeta_{3\pi}$.
The distributions in the $\zeta_{3\pi}$--$\zeta_{2\pi}$ plane for the data and
$K_S \to 3\pi^0$ simulated signal are shown in Fig.~\ref{fig3}. Signal events
are characterized by small values of $\zeta_{3\pi}$ and relatively high
$\zeta_{2\pi}$~\cite{Michal}, which  indicates the choice of the signal region definition (signal box).
In order to improve the quality of the photon selection using $\chi^{2}_{2\pi}$,
we cut on the variable $\Delta E~=~(m_{\Phi}/2 - \sum E_{\gamma_{i}})/\sigma_{E}$
where $\gamma_i$ stands for the i--$th$ photon from  four chosen in the $\chi^{2}_{2\pi}$
estimator and $\sigma_E$ is the appropriate resolution. For $K_S \to 2\pi^0$ decays plus two background
clusters, we expect $\Delta E~\sim~$0, while for $K_S \to 3\pi^0$ $\Delta E~\sim~m_{\pi^0}/\sigma_E$.
To further reject surviving $K_S \to 2\pi^0$ events with split clusters, we cut on the minimal
distance between centroids of reconstructed clusters, $R_{min}$, considering that the distance
between split clusters is on average smaller than the distance between clusters originating
from $\gamma$'s of $K_S \to 3\pi^0$ decay. Distributions of these two discriminant variables are presented
in Fig.~\ref{fig5}.
\begin{figure}
\begin{center}
\includegraphics[width=0.49\textwidth]{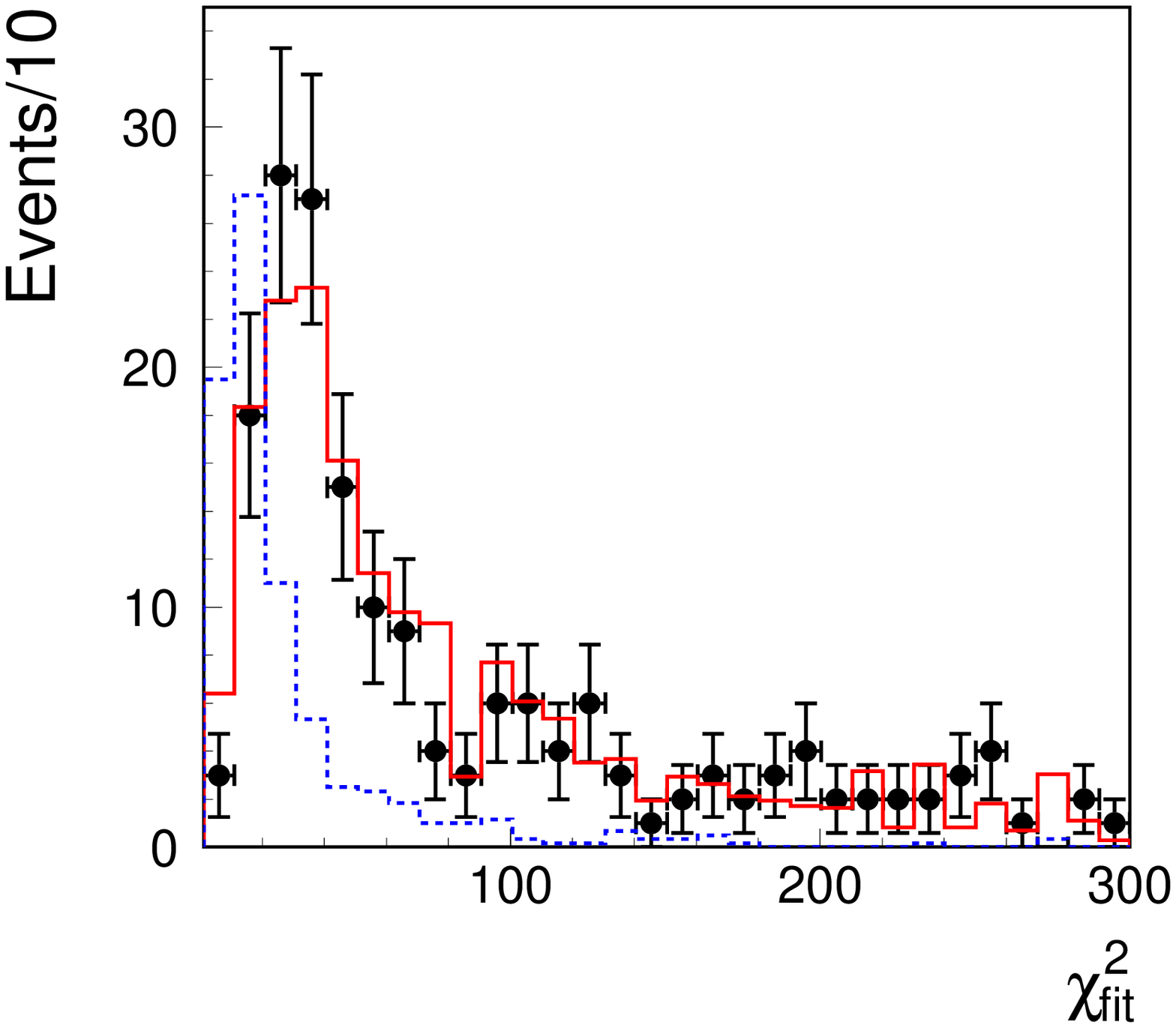}
\includegraphics[width=0.49\textwidth]{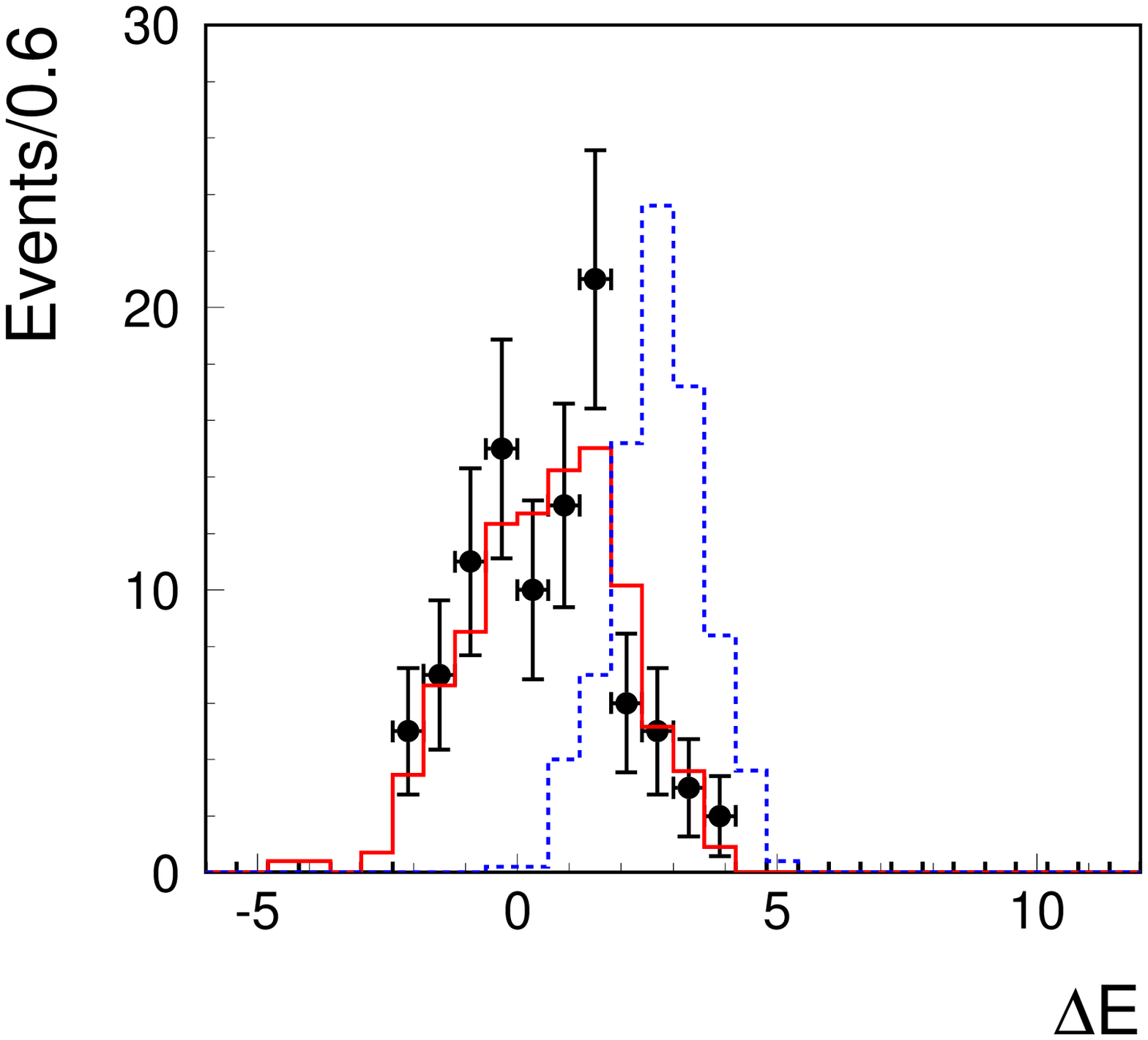}
\end{center}
\caption{Distributions of $\chi^2_{fit}$ for six--photon events in the signal box (left)
and $\Delta E$ for six--photon events in the signal box applying the $\chi^2_{fit} <$~57.2 cut (right).
Black points are data, background simulation is the red histogram.
The dashed histogram represents simulated $K_S \to 3\pi^0$ events.}
\label{fig6}
\end{figure}
Before counting of the $K_S \to 3\pi^0$ candidate events, we have optimized the event selection
in order to reduce the background as strongly as possible while keeping high signal efficiency.
Cuts on the discriminant variables have been
refined minimizing $f_{cut}(\chi^2_{fit},\zeta_{2\pi},\zeta_{3\pi},\Delta E,R_{min}) = N_{up}/\epsilon_{3\pi}$,
where $\epsilon_{3\pi}$ stands for the signal selection efficiency and
$N_{up}$ is the mean upper limit (at 90\% C.L.) on the expected number of signal events. It is calculated
on the basis of the corresponding expected number of background events
$B_{exp} = B_{exp}(\chi^2_{fit},\zeta_{2\pi},\zeta_{3\pi},\Delta E,R_{min})$
from simulation~\cite{silarskiPHD}.
As the result of the optimization we have obtained the following
values for cuts on discriminant variables: $\chi^2_{fit} < 57.2$, $\Delta E > 1.88$ and $R_{min} > 65$~cm.
The signal box is defined as: $4 < \zeta_{2\pi} < 84.9$ and $\zeta_{3\pi} < 5.2$~\cite{Michal}.
Since the expected number of background events was estimated using the Monte Carlo simulations,
we have checked at each stage of the analysis how well the simulation describes the experimental data.
Distributions of $\chi^2_{fit}$, $\Delta E$ and $R_{min}$ variables are presented
in Fig.~\ref{fig6} and Fig.~\ref{fig7} for events in the signal box.
In the right panel of Fig.~\ref{fig7} we present also the $R_{min}$ distribution just before
the last cut $R_{min}>65$~cm marked with a black arrow. One can see a good agreement between simulations
and measured data at every stage of the analysis.
\begin{figure}
\begin{center}
\includegraphics[width=0.49\textwidth]{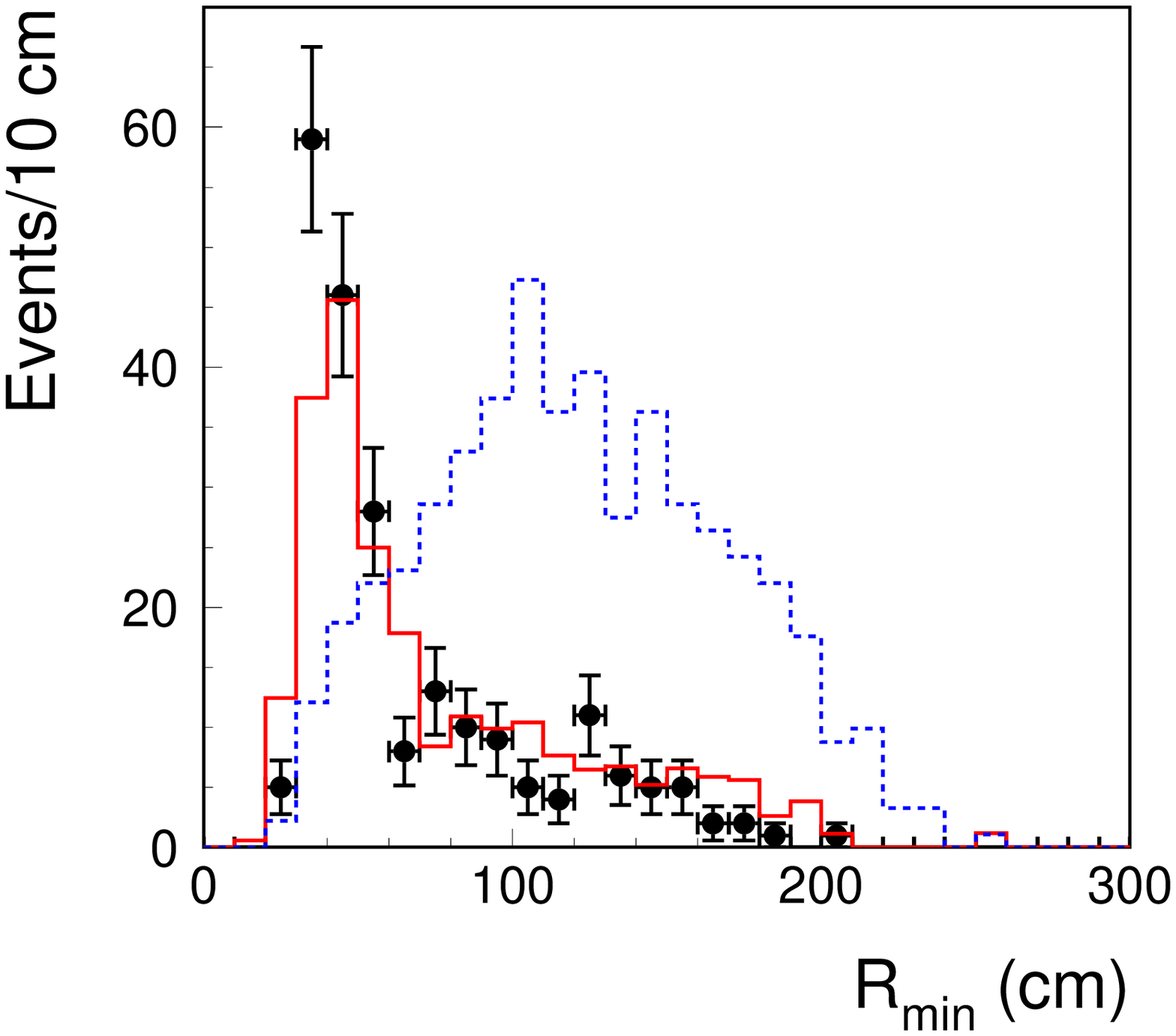}
\includegraphics[width=0.49\textwidth]{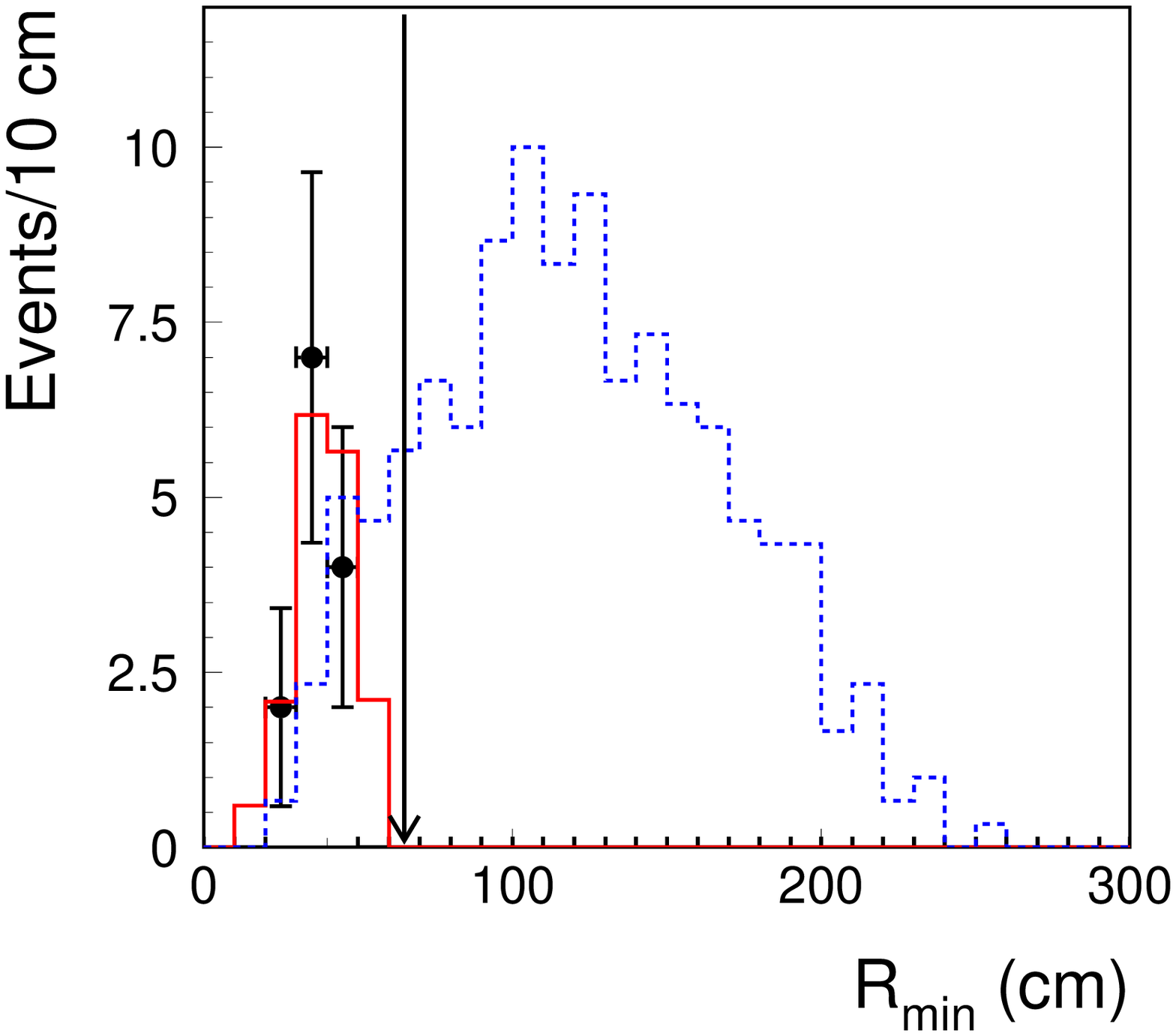}
\end{center}
\caption{Distributions of $R_{min}$ for six--photon events in the signal box applying
the $\chi^2_{fit} <$~57.2 cut (left), and applying $\chi^2_{fit} <$~57.2
and $\Delta E > 1.88$ cuts (right).
Black points are data, background simulation is the red histogram.
The dashed histogram represents simulated $K_S \to 3\pi^0$ events.}
\label{fig7}
\end{figure}
At the end of the analysis we have found zero candidates in data and in the
simulated background sample.\\
As it was mentioned, the $K_S \to 2\pi^0$ normalization sample is selected requiring
four prompt photons. The Monte Carlo simulation shows a negligible (0.1$\%$) amount of
$\phi \to K^+K^-$ background events, thus no further discriminant analysis was required.
In the same tagged sample of $K_S$ mesons we have found $N_{2\pi}~=~(7.533 \pm 0.018) \cdot 10^7$
events. Using simulations we have also determined the $K_S \to 2\pi^{0} \to 4\gamma$
selection efficiency: $\epsilon_{2\pi} = 0.660 \pm 0.002_{stat}$. 
The final number of produced $K_S \to 2\pi^0$ events amounts to:
$ N_{norm} = N_{2\pi}/ \epsilon_{2\pi} = (1.142 \pm 0.005) \cdot 10^8$.
\section{Results}
With cuts defined in the previous section at the end of the analysis we count 0
candidates with 0 background events expected from Monte Carlo simulated with an effective
statistics of two times that of the data. Hence, we have estimated an upper limit on the
$K_S \to 3\pi^0$ branching ratio taking into account systematic uncertainties related to
the number of background events and  to the determination of
the acceptance and selection efficiencies for the signal and normalization samples.
The detailed description of the evaluation of systematic uncertainties can be found
in Ref.~\cite{silarskiPHD, Michal}, here we would like only to stress that they are less than 5\%
and are negligible in the calculation of the limit.\\
In the conservative assumption of no background, we estimate an upper
limit on the $K_S \to 3\pi^0$ branching ratio at 90\% confidence level:
\begin{equation}
\nonumber
BR(K_S \to 3 \pi^0) \leq 2.6 \cdot 10^{-8}~,
\nonumber
\end{equation}
which corresponds to an improvement by a factor of about 5 compared to our
previous search~\cite{Matteo}.
This result can be translated into a limit on $|\eta_{000}|$: 
\begin{equation}
\nonumber
|\eta_{000}| = |{A(K_S \to 3\pi^0)}/{A(K_L \to 3\pi^0)}| \leq 0.0088\;\;\; {\rm at}\;\; \; 90\%\;\;\; {\rm C.L.}
\nonumber
\end{equation}
\section{Summary and outlook}
As a result of the KLOE data set analysis, gathered in the 2004--2005
data taking period, we have set the new upper limit for the $K_S \to 3\pi^0$
branching ratio at the 90$\%$ confidence level,
which is almost five times lower than the latest published result~\cite{Matteo}.
However, the search for the $K_S \to 3\pi^0$ decay will be continued
by the KLOE--2 collaboration, which is continuing
and extending the physics program of its predecessor~\cite{AmelinoCamelia}. 
For the forthcoming run the KLOE performance have been improved by adding new
subdetector systems: the tagger system for the $\gamma\gamma$ physics,
the Inner Tracker based on the Cylindrical GEM technology and two calorimeters
in the final focusing region~\cite{Moricciani:2012zza}. These new calorimeters
will increase the acceptance of the detector, while the new inner detector for
the determination of the $K_S$ vertex will significantly
reduce the contribution of the background processes involving charged particles.
Increasing the statistics and acceptance of the detector while significantly reducing
the background gives a realistic chance to observe the $K_S \to 3\pi^0$ decay for
the first time in the near future.
%
\section*{Acknowledgments}
We acknowledge the support of the European Community-Research
Infrastructure Integrating Activity `Study of Strongly Interacting Matter'
(acronym HadronPhysics2, Grant Agreement n. 227431) under the Seventh Framework
Programme of EU.
This work was supported also in part by the EU Integrated Infrastructure
Initiative Hadron Physics 
Project under contract number RII3-CT- 2004-506078; by the European
Commission under the 7th 
Framework Programme through the `Research Infrastructures' action of
the `Capacities' Programme, Call: 
FP7-INFRASTRUCTURES-2008-1, Grant Agreement No. 283286; by the Polish
National Science Centre through the 
Grants No. 0469/B/H03/2009/37, 0309/B/H03/2011/40, 
DEC-2011/03/N/\\ST2/02641, 2011/01/D/ST2/00748, 2011/03/N/ST2/02652,
2011/03/N/ST2/02641 and by the Foundation for Polish Science through the MPD programme 
and the project HOMING PLUS BIS/2011-4/3.


\section*{References}
\begin{thebibliography}{9}

\bibitem{silarskiPHD}
Silarski M 2012 Search for the CP symmetry violation in the decays of $K_S$
mesons using the KLOE detector \textit{Preprint} arXiv:1302.4427 [hep-ex]

\bibitem{pdg2012}
Beringer J {\it et al.} [Particle Data Group Collaboration] 2012
\textit{Phys.\ Rev.} D \textbf{86} 010001

\bibitem{Matteo}
Ambrosino F {\it et al.} 2005 \textit{Phys.\ Lett.} B \textbf{619} 61

\bibitem{MPaver}
 Maiani L and Paver N 1995 \textit{The Second DA$\Phi$NE Physics Handbook} ed Maiani L, Pancheri G and
 Paver N (Frascati: Frascati Phys. Ser.) p 51

\bibitem{Michal}
Babusci D {\it et al.}  [KLOE-2 Collaboration],
A new limit on the CP violating decay $K_S \to 3\pi^0$ with the KLOE experiment
  \textit{Preprint} arXiv:1301.7623 [hep-ex]
  
\bibitem{kloe2008}
Bossi F {\it et al.} 2008 \textit{Riv.\ Nuovo Cim.} \textbf{31} 531

\bibitem{DCH}
Adinolfi M {\it et al.} 2002 \textit{Nucl.\ Inst.\ and Meth.} A \textbf{488} 51

\bibitem{EMC}
Adinolfi M {\it et al.} 2002 \textit{Nucl. Inst. and Meth.} A \textbf{482} 364

\bibitem{TRG}
Adinolfi M {\it et al.} 2002 \textit{Nucl.\ Inst.\ and Meth.} A \textbf{492} 134

\bibitem{NIMOffline}
Ambrosino F {\it et al.} 2004 \textit{Nucl.\ Inst.\ and Meth.} A \textbf{534} 403

\bibitem{AmelinoCamelia}
Amelino-Camelia G {\it et al.} 2010
  \textit{Eur.\ Phys.\ J.} C \textbf{68} 619

\bibitem{Moricciani:2012zza}
Moricciani D 2011 The KLOE-2 detector upgrade at DA$\Phi$NE
  \textit{PoS} {\bf EPS--HEP2011} 198
\end{thebibliography}
\end{document}